\documentclass[notoc,paper]{JHEP}
\usepackage{amssymb}
\usepackage{cite}
\usepackage{graphicx}
\newcommand{\nco}{\newcommand}
\def\phm{\phantom{-}}
\nco{\half}{\frac{1}{2}}
\nco{\shalf}{\ensuremath{\textstyle \frac{1}{2}}}
\newcommand{\ev}{\ensuremath{\mathrm{\; eV}}}
\newcommand{\mev}{\ensuremath{\mathrm{\; MeV}}}

\title{Oscillation induced neutrino asymmetry growth in
       the early Universe}

\author{Kimmo Kainulainen \\ NORDITA \\Blegdamsvej 17
        \\ DK-2100 Copenhagen \O \\Denmark
        \\e-mail: \email{kainulai@nordita.dk}}

\author{Antti Sorri \\ Department of Physical Sciences
        \\ University of Helsinki
        \\ P.O. Box 64 \\ FIN-00014 University of Helsinki
        \\ e-mail: \email{antti.sorri@helsinki.fi}}

\abstract{We study the dynamics of active-sterile neutrino
oscillations in the early universe using full momentum-dependent
quantum-kinetic equations. These equations are too complicated to
allow for an analytical treatment, and numerical solution is greatly
complicated due to very pronounced and narrow structures in the
momentum variable introduced by resonances. Here we introduce a
novel dynamical discretization of the momentum variable which
overcomes this problem. As a result we can follow the evolution
of neutrino ensemble accurately well into the stable growing phase.
Our results confirm the existence of a ``chaotic region" of mixing
parameters, for which the final sign of the asymmetry, and hence
the SBBN prediction of $^4$He-abundance cannot be accurately
determined.}

\preprint{NORDITA 2001/76 HE}

\begin{document}

\section{Introduction}

Recent observations from atmospheric and solar neutrinos from Super
Kamiokande (SK)\cite{SK1} and Sudbury Neutrino Observatory (SNO)
\cite{SNO} prefer conventional, mainly active-active mixings between
ordinary electron, muon and tau neutrinos as the solutions to the
observed flux deficits. However, while pure active-sterile solutions
are disfavoured, sizable admixtures of sterile states in the observed
fluxes are still allowed~\cite{smirnov}. New sterile states are also
needed in order to explain the LSND anomaly~\cite{LosAl} by neutrino
physics. Moreover, in a recent analysis of the so called
``2+2''-models\footnote{Alternative ``3+1'' models fit less well
with short baseline reactor experiments~\cite{3p1schemes}.}
for four-neutrino mixing, the solutions with nonzero sterile
neutrino components were found to provide best global fits to the
solar, atmospheric and reactor data~\cite{stfluxes}. Corresponding
limits on active-sterile mixings are quite generous, for example
\begin{eqnarray}
\sin^2\theta_{\mu s} &\lesssim 0.48& \qquad \rm (Atmospheric) \nonumber\\
\sin^2\theta_{e s}   &\lesssim 0.72& \qquad \rm (Solar \;\; LMA),
\label{labfluxes}
\end{eqnarray}
where LMA refers to the Large Mixing Angle solution for the solar
neutrino deficit.

Active-sterile mixing, if realized, would have several interesting
effects in astrophysical settings~\cite{snova} and in particular for
the evolution of the early universe
\cite{dol,ektdo,ekm,ektBig,ektat,ekmL,old,ftv,shi,eks1,fSc,Sorri,shifuNc}.
In particular, sterile states could be brought into thermal equilibrium
by mixing before nucleosynthesis, so that the resulting anomalous increase
in the expansion rate of the universe would lead to overproduction of helium
in disagreement with the observations~\cite{dol,ektdo,ekm,ektBig,ektat,old}.
Excluding the parameter sets leading to equilibration provides useful bounds
on active-sterile mixing. Indeed, from results of ref.~\cite{ektBig} one
can infer that the sterile components in mass eigenstates responsible
for atmospheric anomaly and LMA are constrained by
\begin{eqnarray}
\sin^2\theta_{\mu s} &\lesssim 0.013& \qquad \rm (Atmospheric) \nonumber\\
\sin^2\theta_{e s}   &\lesssim 0.026& \qquad \rm (Solar \;\; LMA)
\label{cosmofluxes}
\end{eqnarray}
where we used a SBBN-limit of $N_\nu \lesssim 3.4$ for the number of
effective neutrino degrees of freedom~\cite{ektBig,recent}. These
numbers correspond to the ``light" case where the sterile component is
the heavier of the mixing states; in the opposite, ``dark" case, the
corresponding limits are even stronger by one to two orders of magnitude.
In either case the cosmological constraints are more stringent than
those obtained in terrestrial laboratories.

While quite generic, the cosmological bounds (\ref{cosmofluxes})
do depend on some simple prior assumptions, the most important of
which is that the primordial lepton asymmetry is not anomalously
large\cite{ekmL}.  It may be possible to
circumvent them in more complicated mixing schemes involving more
sterile states (for a recent discussion, see~\cite{dibSc}). In a
particular attempt it was shown in ref.~\cite{ftv} that for a large
negative $\delta m^2$ and a small enough mixing angle (so that
the equilibration bounds of~\cite{ektBig} can be avoided), say on
the $\nu_\tau-\nu_{s'}$-sector, resonant oscillations may trigger
a rapid growth of the tau lepton asymmetry. This asymmetry could then
become very large, $L_\tau \sim {\cal O}(1)$ and change the natural
prior condition of no significant initial lepton asymmetry for a
mixing in any other sectors.  Indeed, if created early enough, such
an asymmetry could suppress $\nu_\mu-\nu_s$-oscillations~\cite{fvL1,ekmL},
obviating the bounds (\ref{cosmofluxes}) for $\nu_\mu-\nu_s$-mixing.
However, as this scenario requires that the sterile state is the
{\em lighter} one (resonance condition), and that the mass splitting
$|\delta m^2|$ is large (to create $L_{\tau}$ early enough), it is
essentially excluded by the recent solar and atmospheric neutrinos
combined with the direct constraint on the electron neutrino
mass~\cite{mainz}.

Even with the mechanism of ref.~\cite{ftv} by and large excluded,
there can be important effects due to $L$-growth with smaller
$|\delta m^2|$.  For example, a large homogeneous electron neutrino
asymmetry $L_{\nu_e}$ would directly influence the nucleosynthesis
prediction for the helium-4 abundance through the reactions
\begin{eqnarray}
 n + e^+  &\leftrightarrow & p + \bar \nu_e \nonumber \\
 p + e^-  &\leftrightarrow & n + \nu_e \;,
\end{eqnarray}
which keep the neutron-to-proton ratio in equilibrium at early times.
Changing the electron neutrino abundance could tilt the balance of these
reactions with the possibility of either increasing $Y_{He}$ (negative
$L_{\nu_e}$), and hence strengthening the bounds (\ref{cosmofluxes}), or
decreasing $Y_{He}$ (positive $L_{\nu_e}$), leading to weakening of the
bounds (\ref{cosmofluxes}).

The physics involved with the resonant growth of the asymmetry is very
complicated because of several vastly different physical scales both
in the temporal direction and in momentum variable, which effect the
evolution of the ensemble in an essential way. The relevant QKE's are
nonlinear and strongly coupled internally via the asymmetry term, and
therefore no useful analytical approximation exists for the
problem\footnote{
Some confusion in the field was caused by a recent analytical
treatment~\cite{dolHan}, whose results contradicted earlier numerical
results~\cite{shi,eks1,fSc}. Disagreement was eventually clarified in
favour of numerical work in~\cite{DiBetal}.}.
Numerical solution is also greatly complicated by the various different
physical scales, and
while the phenomenon of asymmetry growth is fairly well established
by now, the determinacy of the final {\em sign} of the asymmetry is
still not well understood. The ambiguity was first observed in~\cite{shi},
and in~\cite{eks1,Sorri} it was shown that this ``chaoticity" occurs in
certain well defined region of mixing parameters.
References~\cite{shi,eks1,Sorri} employed a numerical solution for the
momentum averaged approximation of the QKE's however, and their results
were challenged by ref.~\cite{dolHan}, who claimed that the final
asymmetry is always fully determined by the initial conditions. While
the analysis of ref.~\cite{dolHan} is now discredited~\cite{DiBetal},
it still remains true that no numerically reliable momentum dependent
analysis has so far shown the existence of the chaotic
region.\footnote{
Although ref.~\cite{fdb} found support for chaoticity,
their conclusion is not firm because of the reported loss of the
numerical accuracy of their methods when approaching the potentially
chaotic region.}
It is important to settle the issue because, as mentioned above, large
$L_{\nu_e}$ could significantly alter the helium-4 abundance. Indeed,
if the sign of $L_{\nu_e}$ were found to be chaotic, then $Y_{He}$
could not be reliably determined from BBN~\cite{eks1}.

In this paper we study the dynamics of active-sterile neutrino
oscillations in the early universe using full momentum-dependent
quantum-kinetic equations (in homogeneous space). Our main technical
improvement over the previous works is the introduction of a novel
dynamical discretization method for the momentum variable which
enables us to model accurately the density matrices over the entire
momentum range, including the extremely pronounced and narrow
structures close to resonant momenta.  On physical context, our
results confirm the existence of a "chaotic" region of mixing
parameters, for which the final sign of the asymmetry is not
deterministic. It also confirms the expectation~\cite{eks1,fdb}, that
the size of the chaotic region is somewhat smaller than indicated by
the momentum averaged code~\cite{eks1}. These results can be seen as
giving more strength on the cosmological constraints on neutrino
mixing parameters.

The paper is organized as follows. In section 2 we set up the quantum
kinetic equations for the problem. In section 3 we introduce the novel
dynamically adjusted discretization of the momentum variable and in
section 4 we set up our kinetic equations with this parametrization.
The method we develop allows us to have enough resolution to solve the
problem with relatively small number of momentum bins (we use at most
400 bins).  In section 5 we display our numerical results for the key
variables driving the oscillation and discuss the numerical stability
of our solutions.  Finally, in section 6 we present our results on the
asymmetry growth and oscillations as well as the interpretation of
the physical consequences, and the section 7 contains a summary and
outlook.

\section{Quantum Kinetic Equations}

In the early Universe the oscillating neutrinos experience frequent
scatterings which interrupt the coherent evolution of the state and
introduce collisional mixing between different momentum states. The
mathematical formalism which can incorporate all these features has
been developed elsewhere~\cite{stod,ekm,ektBig,sigl,TMcK}, and it takes
the form of the quantum kinetic equations for the reduced density
matrices for the neutrino and antineutrino ensembles. We parametrize
the density matrices by the Bloch vector presentation~\cite{ektBig}:
\begin{equation}
\rho_{\nu}      \equiv \frac{1}{2} f_0 (P_0
          + {\bf P} \cdot {\bf\sigma}) \; , \qquad
\rho_{\bar \nu} \equiv \frac{1}{2} f_0 (\bar P_0
          + {\bf \bar P} \cdot {\bf \sigma}),
\label{rho}
\end{equation}
where $\mathbf{\sigma}$ are the Pauli spin matrices and
$f_0 = (1+\exp(p/T))^{-1}$ is the Fermi-Dirac distribution
function without chemical potential. Then to the the lowest
order in the interaction~\cite{stod,ekm,sigl,TMcK}, the
QKE's for the density matrix take the form
\begin{eqnarray}
d_t P_0
     &=& \phm \frac{\Gamma}{f_0} \left[ f_{eq}
    - \rho_{\alpha \alpha} \right] \nonumber \\
d_t P_x
     &=& -    V_z P_y - D P_x \nonumber \\
d_t P_y
     &=& \phm V_z P_x - V_x P_z - D P_y \nonumber \\
d_t P_z
     &=& \phm V_x P_y + \frac{\Gamma}{f_0} \left[ f_{eq}
                               - \rho_{\alpha \alpha} \right] \, ,
\label{eq:master}
\end{eqnarray}
where $d_t \equiv \partial_t - H p \partial_p$ and $H$ is the
Hubble expansion factor. The rotation vector $\mathbf{V}$ has
the following components ($V_y \equiv 0$):  in $x$-direction
one has only the vacuum contribution
\begin{equation}
    V_x = \frac{\delta m^2}{2 p} \sin 2 \theta_0 \;,
\end{equation}
whereas in the $z$-direction also the matter induced effective
potential contributes:
\begin{equation}
    V_z  = V_0  + V_1  + V_L,
\label{vzeta}
\end{equation}
where
\begin{eqnarray}
  V_0 \phm       &=& - \frac{\delta m^2}{2 p} \cos 2 \theta_0, \\
  V_1^e \phm     &=& - 14 \sqrt{2} \frac{\zeta(4)}{\zeta(3)}
                             \frac{G_F}{M_W^2} N_{\gamma} p \; T
                   \left( 1 + {\textstyle \frac{1}{4}} \cos^2 \theta_W
                   \left[ n_{\nu_e} + n_{\bar \nu_e } \right] \right) \\
  V_1^{\mu,\tau} &=& - \frac{7 \sqrt{2}}{2} \frac{\zeta(4)}{\zeta(3)}
                             \frac{G_F}{M_Z^2} N_{\gamma} p \; T
                        \left[ n_{\nu_{\mu,\tau}}
                        + n_{\bar \nu_{\mu,\tau} } \right] \\
  V_L \phm       &=& \phm \sqrt{2} G_F N_{\gamma} L^{(\alpha)}.
\end{eqnarray}
Here $N_\gamma$ is the photon equilibrium number density,
$n_{\nu_{\alpha,(\bar\alpha)}}$ is the normalized to unity (in
equilibrium) actual neutrino (antineutrino) number density, and the
effective neutrino asymmetries $L^{(\alpha)}$ are given by
\begin{eqnarray}
    L^{(e)}    &=&  \left( \shalf + 2 \sin^2 \theta_W \right) L_e
                  + \left( \shalf - 2 \sin^2 \theta_W \right) L_p
                  - \shalf L_n + 2 L_{\nu_e}
                  + L_{\nu_{\mu}} + L_{\nu_\tau} \\
    L^{(\mu)}  &=&  L^{(e)} - L_e -L_{\nu_e} + L_{\nu_{\mu}} \\
    L^{(\tau)} &=&  L^{(e)} - L_e -L_{\nu_e} + L_{\nu_{\tau}} .
\end{eqnarray}
where $L_f \equiv (n_f - n_{\bar f})N_f/N_\gamma$.

The repopulation term $\Gamma (f_{eq} - \rho_{\alpha \alpha})$ in the
diagonal part of (\ref{eq:master}) is written in the relaxation time
approximation~\cite{aussiet}. The distribution $f_{eq}$ in the
repopulation term is the usual equilibrium Fermi-Dirac distribution
\begin{equation}
    f_{eq} = \frac{1}{1 + e^{\frac{p}{T}-\frac{\mu}{T}}}.
\end{equation}
and the reaction rates are
\begin{equation}
    \Gamma = C_\alpha G_F^2 p T^4
\end{equation}
with $C_e \simeq 1.27$ and $C_{\mu,\tau} \simeq 0.92$\cite{ektBig}.
The damping terms $D P_i$ correspond to the decohering scatterings,
which tend to destroy the coherence of the oscillation. In density
matrix language such terms appear as relaxation terms suppressing
the off-diagonals of the density matrix, or equivalently the
$P_{x,y}$-components of the Bloch vectors. The magnitude of the
decohering terms is just half of the corresponding scattering rate
\begin{equation}
    D = \half \Gamma.
\end{equation}
The equation of motion for anti-neutrinos can be found by substituting
$L^{(\alpha)} = -L^{(\alpha)}$ and $\mu_{\nu} = - \mu_{\nu}$ to the
above equations (the latter condition is true when neutrinos are in
chemical equilibrium).

The repopulation term used above is only an approximation for the
correct elastic collision integral, and unfortunately it breaks the
lepton number conservation. This is not a serious problem however,
and it can be circumvented by introducing an explicit lepton number
conserving evolution equation for $L^{(\alpha)}$.  The appropriate
equation, introduced in~\cite{fdb}, is
\begin{equation}
  d_t L^{(\alpha)}
         = \frac{1}{8\zeta(3)}
           \int\limits_0^\infty dp \;
                    p^2 (V_x P_y - \bar V_x \bar P_y).
\label{L-equation}
\end{equation}
Equation~(\ref{L-equation}) can be derived from the original equations
in the approximation where the $L$-violation due to the the approximate
repopulation term is neglected. So, when using (\ref{L-equation})
to evolve $L$, it is guaranteed that the repopulation term does not
{\em directly} affect the lepton asymmetry generation. The accuracy
of the method can be monitored by comparing the value of $L^{(\alpha)}$
derived from (\ref{L-equation}) and from directly integrating the
neutrino distribution functions. In practice the agreement was always
to be good.

Neutrino and antineutrino sectors are extremely strongly coupled
through the asymmetry $L^{(\alpha)}$ in equations~(\ref{eq:master}). In
order to compute $L^{(\alpha)}$ accurately enough despite the numerical
round-off errors, it is then necessary to avoid computing $L^{(\alpha)}$
through differences of large quantities, such as naively appear in the
equation (\ref{L-equation}).  To this end we will define the ``small"
and ``large" linear combinations of dynamical variables in particle
and antiparticle sector
\begin{equation}
P_i^\pm = P_i \pm \bar P_i .
\label{Pplusminus}
\end{equation}
For convenience we also separate active and sterile sectors
by defining
\begin{eqnarray}
P_a^\pm & = P_0^\pm + P_z^\pm & = 2 \rho_{\alpha \alpha}^\pm/f_0
\label{eq:newvar1}\\
P_s^\pm & = P_0^\pm - P_z^\pm & = 2 \rho_{ss}^\pm/f_0.
\label{eq:newvar2}
\end{eqnarray}
In terms of variables (\ref{eq:newvar1}-\ref{eq:newvar2}) the
equations take the form
\begin{eqnarray}
d_t P_a^\pm & = & \phm  V_x P_y^\pm + \Gamma  \left[2 f_{eq}^\pm/f_0
                      - P_a^\pm \right] ,
\label{aktiivi} \\
d_t P_s^\pm & = &   -   V_x P_y^\pm  ,
\label{steriili} \\
d_t P_x^\pm & = &   -  (V_0 + V_1) P_y^\pm  - V_L P_y^\mp - D P_x^\pm , \\
d_t P_y^\pm & = & \phm (V_0 + V_1) P_x^\pm  + V_L P_x^\mp
                  - \half V_x (P_a^\pm - P_s^\pm)  - D P_y^\pm ,
\label{eq:plusminus}
\end{eqnarray}
where we assumed $\Gamma = \bar \Gamma$ and defined
\begin{equation}
f_{eq}^\pm  =  f_{eq} (p,\mu) \pm f_{eq} (p,-\mu)
\end{equation}
Because $\bar V_x = V_x$, the equation for the asymmetry now reads
simply
\begin{equation}
  d_t L^{(\alpha)}
         = \frac{1}{8\zeta(3)}
           \int\limits_0^\infty dp \; p^2 V_x P_y^-.
\label{L-equation2}
\end{equation}
Finally, due to the particular form of the differential operator
$d_t = \partial_t - Hp\partial_p$, it is possible to reduce our
set of partial differential equations to ordinary differential
equations by changing variables $p \rightarrow x \equiv p/T$, and
$t \rightarrow T$, and correspondingly using the time-temperature
relation
\begin{equation}
 \frac{dT}{dt} = -HT.
\end{equation}
The differential operator then becomes simply
\begin{equation}
 d_t \rightarrow  - \frac{1}{HT} \partial_T.
\end{equation}
Let us stress that our introducing the small and large linear
combinations is an absolutely necessary step for obtaining reliable
solutions to
QKE's~(\ref{aktiivi}-\ref{eq:plusminus},\ref{L-equation2}).
This can best be appreciated from the figure (\ref{fig:rplus}) below,
where we show the energy spectrum in neutrino {\em and} antineutrino
sectors for a representative set of oscillation parameters; the
asymmetry corresponds to the integral over the difference (not even
visible!) of the neutrino and antineutrino distributions displayed.

%
%
\section{Discretization of the momentum variable}

While necessary, introducing parametrization (\ref{Pplusminus}) is
unfortunately not sufficient to tackle the momentum dependent QKE's.
Additional difficulties arise due to structures in the momentum
direction (in variable $x$) and in particular the very narrow ones
introduced by the resonances. In this paragraph we explain step by
step how we discretize the momentum variable such that these
difficulties can be overcome.

First, the momentum variable ranges over the semi-infinite range
$x \in [0,\infty]$, but in practice we will introduce cut-offs to both
ends of spectrum. First, the ultra-relativistic neutrino approximation
breaks down when $x\rightarrow 0$, formally appearing as a singularity
in at $x=0$ in our equations.  Second, it is useful to cut out very
large $x$, in the exponentially suppressed tail of the distribution.
So, we restrict $x$ to a range $x \in [x_{min},x_{max}]$, where in
practice it suffices to use\footnote{Note that even for
$x=10^{-4}$ still $p \gtrsim 1$ keV throughout our computation, so
that the ultrarelativistic approximation always remains valid.}
$x_{min} = 10^{-4}$ and $x_{max} = 100$.

Second, the variable $x$ is not ideal for discretization, because
most of the variation in the spectrum occurs at small $x \lesssim 3$.
We therefore have mapped the variable $x$ to $u(x)$ belonging to
range $[0,1]$ by a transformation
\begin{equation}
  u(x) = \frac{x(1+\epsilon_2 ) - x_{\rm ext} \epsilon_1 }{x+x_{\rm ext}}.
\label{mapping}
\end{equation}
where $\epsilon_1 \equiv k_1(1+\epsilon_2)$ and
$\epsilon_2 \equiv (1+k_1)/(k_2-k_1)$
with $k_1 \equiv x_{\rm min}/x_{\rm ext}$ and
$k_2 \equiv x_{\rm max}/x_{\rm ext}$.
Apart from a small correction due to the cut-off parameters
(\ref{mapping}) is designed so that the extremum point of the thermal
momentum distribution $x_{ext}$ gets mapped to $u \simeq 1/2$.
We will also need the inverse of this function, which is
\begin{equation}
   x(u) = \frac{x_{\rm ext}(u+\epsilon_1)}{1+\epsilon_2-u}.
\label{imapping}
\end{equation}
While clearly a significant improvement for binning the momentum
variable, (\ref{mapping}) is not clever enough to solve the numerical
problems arising from the the pronounced structures around the
resonant momenta. Fortunately, the resonance positions $x_{r_1}$ and
$x_{r_2}$ can be solved analytically from the conditions $V_z = 0$
(neutrinos) and $\bar V_z = 0$ (antineutrinos):
\begin{eqnarray}
    x_{r_1} &=& \sqrt{\varphi^2  + \chi} - \varphi
         \qquad \bar \nu{\rm -resonance}
    \label{eq:rescond1} \\
    x_{r_2} &=& \sqrt{\varphi^2  + \chi} + \varphi
         \qquad \nu{\rm -resonance}
    \label{eq:rescond2}
\end{eqnarray}
where in the case of $\nu_\mu$ and $\nu_\tau$ neutrinos
\begin{eqnarray}
    \varphi &=& \frac{\zeta(3) M_Z^2 L^{(\alpha)}}{14 \zeta(4) T^2}
    \approx 6.60 \times 10^8 L^{(\alpha)} T^{-2}_\mev  \\
    \chi    &=& \frac{\zeta(3) M_Z^2 |\delta m^2|
        \cos 2\theta_0}{14 \sqrt{2} \zeta(4)G_F N_{\gamma} T^3}
    \approx 1.64 \times 10^{8} |\delta m^2|_{\ev^2}
        \cos 2 \theta_0 T^{-6}_\mev.
\end{eqnarray}
Using (\ref{mapping}) one easily finds the corresponding resonant
values in the $u$-variable.

The resonance widths can be estimated from the distance over which
the matter mixing angle drops to a some fraction of its maximum value
of unity.  In this way one finds that the relative resonance width is
proportional to the vacuum mixing angle:
\begin{equation}
\frac{\Delta x}{x} \sim \tan 2\theta_0.
\label{reswidth}
\end{equation}
From (\ref{reswidth}) it is clear that for small mixing angles the
structures near resonances can only be resolved by an extremely
finely spaced momentum grid. For example\footnote{One is driven
to such small values of vacuum mixing to find acceptable phenomenology
for the asymmetry growth~\cite{ftv,fvSc}.}
with $\sin^22\theta_0 = 10^{-6}$, equation (\ref{reswidth}) predicts
structures in the scale $\Delta x/x \simeq 10^{-3}$. Assuming then
that at least 50 points are needed to describe the resonant region
accurately, one sees that a linear binning of the $x$-variable with
$x_{max}=100$ would require $5 \times 10^6$ grid points! Solving such
problem is clearly beyond capabilities of any computer we have today.

Our mapping (\ref{mapping}) helps the situation by roughly a factor
of hundred, but the remaining problem would still be way too large.
We therefore need to find a second transformation $u \rightarrow u(v)$
that would redistribute the points such that more grid points are
clustered around the resonances in the physical momentum variable.
There are obviously many ways to implement such a transformation. We
found it most convenient to construct it by using polynomials of the
form $a + b (v-v_r)^3$. Because we have two resonance points the
mapping function is a little complicated:
\begin{equation}
u(v) = \alpha v + (a + b(v-v_{r_1})^3)\, \theta(v_c-v)
        + (c + d(v-v_{r_2})^3)\, \theta(v-v_c),
\label{eq:v2u}
\end{equation}
where the two separate mappings around resonant values $v_{r_1}$
and $v_{r_2}$ are joined smoothly at the geometric mean of the
resonances $v_c \equiv \frac{1}{2}(v_{r_1} + v_{r_2})$. The linear
term $\alpha v$ was added in order to have a control how steep
is the distribution function of the grid points.

The mapping (\ref{eq:v2u}) has six free parameters $a$, $b$, $c$,
$d$, $v_{r_1}$ and $v_{r_2}$, which will be determined from boundary
conditions at $v=0$ and $v=1$ and continuity conditions at $v=v_c$.
First, the constants $a$, $b$, $c$ and $d$ are fixed by the boundary
conditions
\begin{eqnarray}
    u(0)       &=& 0, \\
    u(v_{r_1}) &=& u_{r_1}, \\
    u(v_{r_2}) &=& u_{r_2}, \\
    u(1)       &=& 1,
\end{eqnarray}
which results in
\begin{eqnarray}
    a &=& u_{r_1} - \alpha v_{r_1} \\
    b &=& \frac{u_{r_1}}{v_{r_1}^3} - \frac{\alpha}{v_{r_1}^2} \\
    c &=& u_{r_2} - \alpha v_{r_2} \\
    d &=& \frac{1-u_{r_2}}{(1-v_{r_2})^3} - \frac{\alpha}{(1-v_{r_2})^2}.
\end{eqnarray}
Note that $u_{r_i}$ are not unknowns, but can be solved analytically
from the resonance conditions (\ref{eq:rescond1}-\ref{eq:rescond2}) and
the inverse mapping (\ref{imapping}). Finally, the continuity conditions
\begin{eqnarray}
 \lim_{v \rightarrow v_c +} & u(v) &
                = \lim_{v \rightarrow v_c -} u(v) \\
 \lim_{v \rightarrow v_c +} & \partial_v u(v) &
                = \lim_{v \rightarrow v_c -} \partial_v u(v).
\end{eqnarray}
provide implicit equations for the remaining two unknowns $v_{r_i}$:
\begin{eqnarray}
        a + \frac{1}{8} b (v_{r_2} - v_{r_1})^3
    &=& c + \frac{1}{8} d (v_{r_1} - v_{r_2})^3
    \label{eq:match.cond1} \\
    b &=& d.
    \label{eq:match.cond2}
\end{eqnarray}
Unfortunately these are sixth order algebraic equations and can only
be solved by iterative numerical algorithms.

\begin{figure}
\centering
\includegraphics[height=7.5cm]{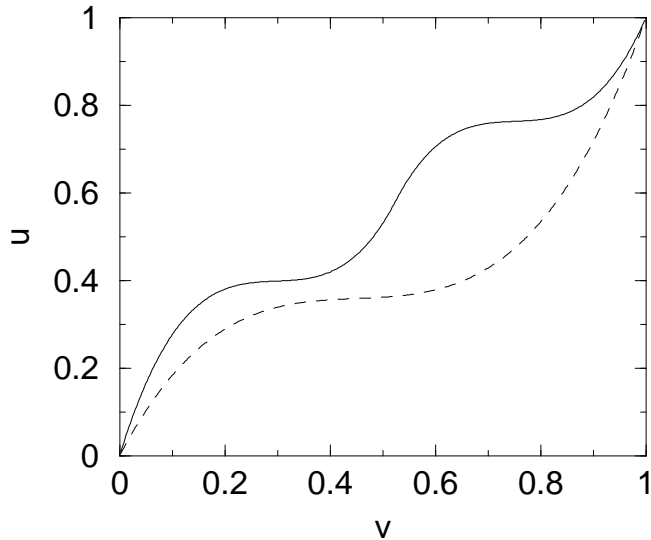}
    \caption{The mapping $u=u(v)$ is shown with  $L^{(\alpha)} =
    4 \times 10^{-7}$, $T=11 \mev$ (solid line) and with
    $L^{(\alpha)} = 10^{-10}$, $T=15 \mev$ (dashed line).
    Mixing parameters are $\delta m^2 = -0.1 \ev^2$
    and $\sin^2 2 \theta_0 = 10^{-9}$. }
    \label{fig:param}
\end{figure}

\begin{figure}
\centering
\includegraphics[height=7.5cm]{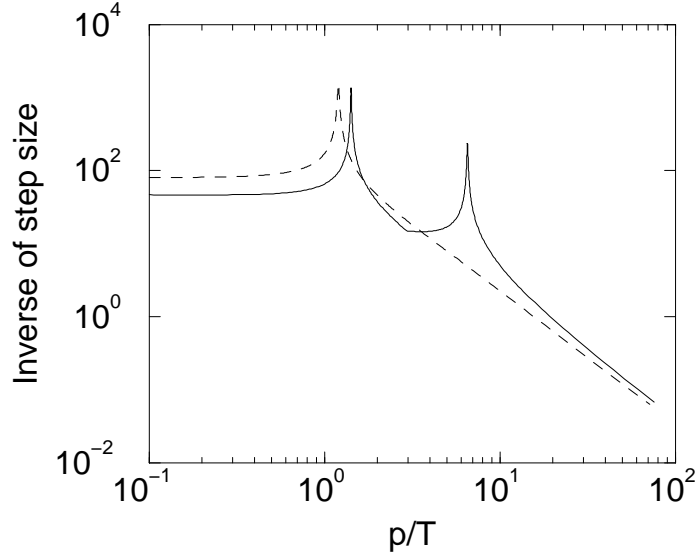}
    \caption{Density of points (defined as the inverse of the step
    size) in the physical momentum variable $x$ for the transforms
    shown in figure~(\ref{fig:param}).}
    \label{fig:param2}
\end{figure}

The novel feature of the above change of variables is that it
automatically follows the resonances, placing majority of the grid
points dynamically in their vicinity. As an alternative to our approach,
one could think of using library routines with automatic grids to take
care of this distribution.  However, the resonance positions evolve as
a function of expansion scale of the universe and in particular as a
function of $L^{(\alpha)}$.  High point densities are thus needed at
different momenta at different times, and this movement can be fast.
None of the library routines we tried came close to being able to
cope with these requirements and failed completely to solve the
problem.  With our parametrization on the other hand, the numerical
program can be written for a fixed grid in the variable $v$, and the
redistribution of points is provided by the dynamical equations
themselves.

In Fig.~(\ref{fig:param}) we show the transformations $u \rightarrow v$
when $L^{(\alpha)}$ is almost zero (dotted line), and when $L^{(\alpha)}$
has grown to a substantial value (solid line). In the former case one
cannot distinguish the two resonances by eye, whereas in the second
case the structures have become very distinct. Fig.~(\ref{fig:param2})
shows the density of points as a function of $x$ (defined as the
inverse of separation of grid points in the $x$-variable).

\section{Changes in the QKE's due to the parametrization}

The changes of variables (\ref{mapping}) and (\ref{eq:v2u})
affect the partial derivatives appearing in
Eqns.~(\ref{aktiivi}-\ref{eq:plusminus},\ref{L-equation2}).
These equations have been written assuming the evolution of
$\rho$ and $L$ is along constant $x$ contours. However, we have
now fixed our grid in the variable $v$, and hence need to find the
evolution equations with $T$ and $v$ as the independent variables.
These are easily found from the old ones by the chain rule
partial differentiation
\begin{eqnarray}
\Big( \frac{\partial \rho\left(T,x(T,v)\right)}{\partial T} \Big)_v
    &=&
        \Big( \frac{\partial \rho}{\partial T} \Big)_x
    +   \Big(\frac{\partial x}{ \partial T} \Big)_v \,
        \Big( \frac{\partial \rho}{\partial x} \Big)_T \nonumber \\
    &=&
        \Big( \frac{\partial \rho}{\partial T} \Big)_x
    +   \Big( \frac{\partial x}{\partial T } \Big)_v
        \Big( \frac{\partial u}{\partial x} \Big)_T
        \Big( \frac{\partial v}{\partial u} \Big)_T \,
        \Big( \frac{\partial \rho}{\partial v} \Big)_T \nonumber \\
    &=&
        \Big( \frac{\partial \rho}{\partial T} \Big)_x
    +   \Big( \frac{\partial u}{\partial T } \Big)_v
        \Big( \frac{\partial v}{\partial u} \Big)_T \,
        \Big( \frac{\partial \rho}{\partial v} \Big)_T \,.
    \label{eq:master.changed}
\end{eqnarray}
In the last equality we used the fact that $u=u(x)$ only. The
differentials $(\partial_T \rho)_x$ are of course just given by the l.h.s.\
of equations (\ref{eq:plusminus}), and the last coefficient $(\partial_v \rho)_T$
is computed numerically as a part of the system of the partial differential
equations. To be precise, we are using the central derivative formula
\begin{equation}
\left( \frac{\partial \rho}{\partial v} \right)_T
                  = \frac{\rho(v+h) - \rho(v-h)}{2h}.
\end{equation}
Of the remaining differentials $(\partial_v u)_T$ is  also easily solved
from the parametrization Eq.~(\ref{eq:v2u}:
\begin{equation}
    \left( \partial_v u \right)_T
    =   \frac{1}{\alpha + 3[ b(v-v_{r_1})^2 \, \theta(v_c-v)
    +   d(v-v_{r_2})^2 \, \theta(v-v_c)]}.
    \label{eq:dvdu}
\end{equation}
However, solving the remaining differential $\left(\partial_T u \right)_v $
is somewhat more complicated. Using equation~(\ref{eq:v2u}) one can write
\begin{eqnarray}
    \left(\partial_T u \right)_v
    &=& \phm \left[ \partial_T a + \partial_T b (v-v_{r_1})^3
    -   3b (v-v_{r_1})^2 \partial_T v_{r_1} \right] \theta(v_c - v)
    \nonumber \\ &&
    +   \left[ \partial_T c + \partial_T d (v-v_{r_2})^3
    -   3d (v-v_{r_2})^2 \partial_T v_{r_2} \right] \theta(v-v_c),
\end{eqnarray}
where
\begin{eqnarray}
    \partial_T a &=& \partial_T u_{r_1} - \alpha \partial_T v_{r_1} \\
    \partial_T b &=& \frac{1}{v_{r_1}^3}
                   \left( \partial_T u_{r_1}
                    +\Big(2 \alpha - \frac{3 u_{r_1}}{v_{r_1}}\Big)
                                        \partial_T v_{r_1} \right) \\
    \partial_T c &=& \partial_T u_{r_2} - \alpha \partial_T v_{r_2}\\
    \partial_T d &=& - \frac{1}{(1 - v_{r_2})^3}
                   \left(\partial_T u_{r_2}
                   + \Big(2 \alpha - 3\frac{1-u_{r_2}}{1-v_{r_2}}\Big)
                                         \partial_T v_{r_2} \right).
\end{eqnarray}
Here the only unknowns are the temperature derivatives of the resonances
in the $v$-variable $\partial_T v_{r_i}$.  These can be solved from our
original matching conditions (\ref{eq:match.cond1}) and
(\ref{eq:match.cond2}), which can formally be written as
\begin{eqnarray}
    f_1 \left(v_{r_1}(T), v_{r_2}(T), T \right) &=& 0
\label{ekuf1} \\
    f_2 \left(v_{r_1}(T), v_{r_2}(T), T \right) &=& 0.
\label{ekuf2}
\end{eqnarray}
Differentiating equations (\ref{ekuf1}-\ref{ekuf2}) with respect to $T$
results in a {\em linear} set of equations for $\partial_T v_{r_i}$, which
after a little algebra can be brought to the form
\begin{equation}
    \partial_T v_{r_i} = U_{ij} \partial_T u_{r_j},
\label{partialTv}
\end{equation}
where the matrix $U_{ij}$ depends both on $u_{r_i}$ and
$v_{r_i}$.  Note that despite the apparent complexity of our
parametrization, we have been able to reduce everything except solving
for the resonance parameters $v_{r_i}$ to a simple linear algebra. Solving
$v_{r_i}$ from the matching equations at each time step would be very
time consuming however. This is so in particular because we need to
solve them with very high accuracy in order for not to lose the
continuity of the matching at $v=v_c$ and to avoid inducing random
numerical errors into the equations.  Fortunately, the equation
(\ref{partialTv}) itself offers the way out of this problem. Namely,
we only need to solve $v_{r_i}$ from the matching conditions
(\ref{eq:match.cond1}) and (\ref{eq:match.cond2}) once, in the
beginning of the iteration, as their value can later be {\em evolved}
by equations (\ref{partialTv}) as the solution proceeds. In this way
even solving for the parametrization from $v$ to $x$ becomes part of
the dynamical equations.  The accumulation of numerical error can be
easily traced by solving the matching conditions independently for
some temperatures. In practice the errors do not accumulate at all
and it turns out that using the differential equation is by far the
superior method for solving $v_{r_i}$ in comparison to using the
matching conditions at each temperature step.

\section{Numerical Results}

\begin{figure}
\centering
\includegraphics[height=7.5cm]{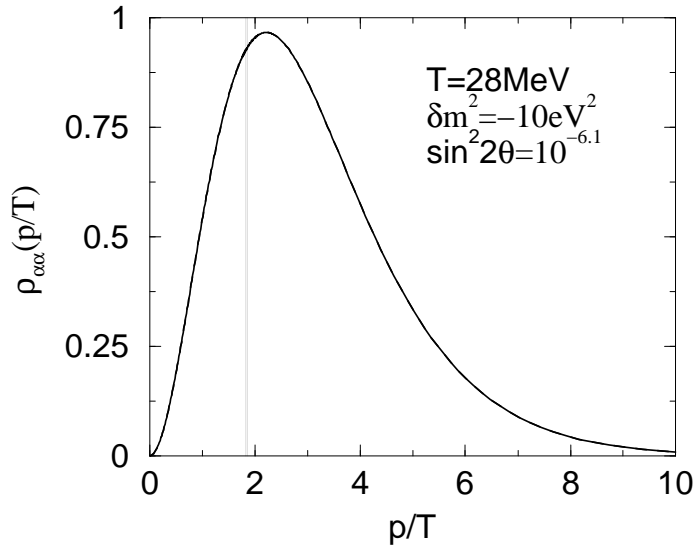}
    \caption{Example of neutrino ($\rho_{\alpha \alpha}$) and
     antineutrino ($\rho_{\bar \alpha \bar \alpha}$) spectra as a
     function of $p/T$. Lines fall on top of each others. Gray
     line marks the resonance momenta.}
    \label{fig:rplus}
\end{figure}

We have solved the evolution equations numerically for a range of
representative parameter sets. In figure (\ref{fig:rplus}) we show the
neutrino and antineutrino distribution functions in a relatively early
stage in the evolution for $\delta m^2 = -10 \; \rm eV^2$ and
$\sin^22\theta_0=10^{-6.1}$. The distributions show a very nice thermal
shape, and no visible difference between particle and antiparticle
sectors can be seen. This is so despite the fact that at the temperature
at which the distributions were plotted, the momenta $p \simeq 1.8T$ are
currently resonant (indicated by the gray line in the plot); the
amplitude of resonant structure is simply too small to be seen in
the scale of the ``large variables".

\begin{figure}
\centering
\includegraphics[height=7.5cm]{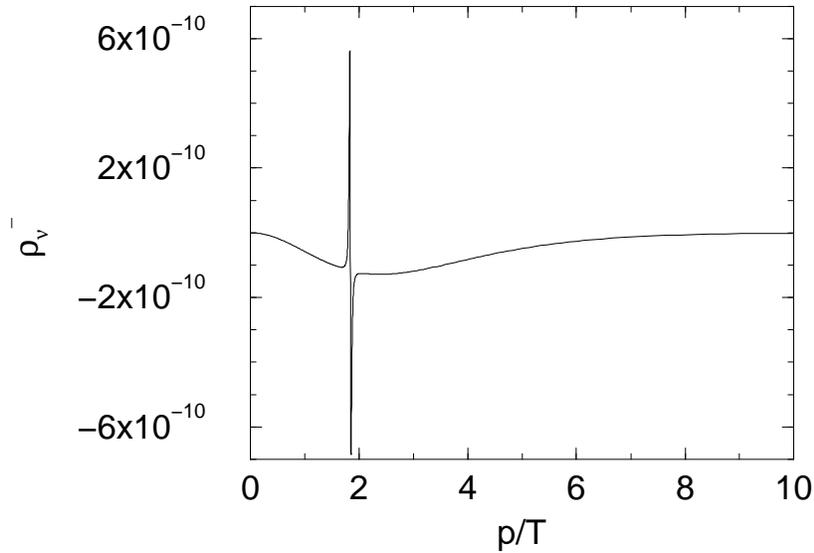}
    \caption{Difference of neutrino and antineutrino densities
    $\rho_{\alpha \alpha} - \rho_{\bar \alpha \bar \alpha}$
    as a function of $p/T$ for the same parameters as in
    figure~{\ref{fig:rplus}}.}
    \label{fig:llx}
\end{figure}

\begin{figure}
\centering
\includegraphics[height=7.5cm]{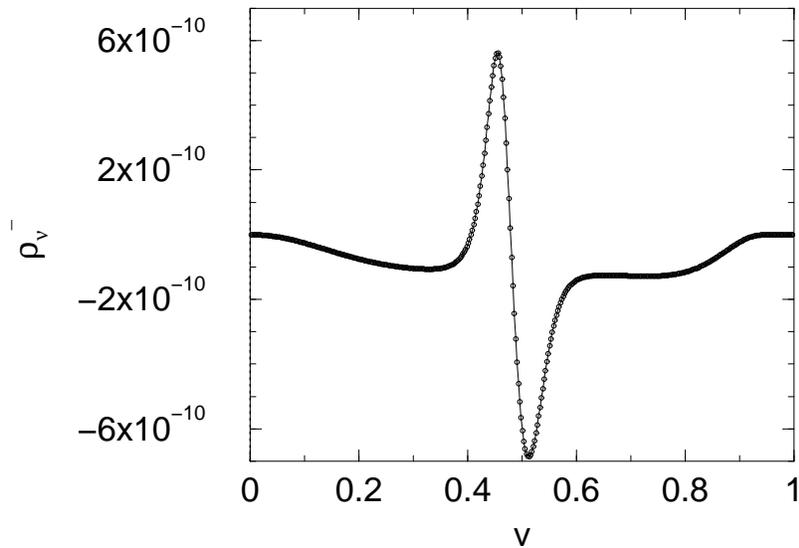}
    \caption{Same as figure~(\ref{fig:llx}), but shown against
    variable $v$.}
    \label{fig:llv}
\end{figure}

The resonance does alter the asymmetry spectrum however. In figure
(\ref{fig:llx}) we show the distribution of the difference of neutrino and
antineutrino densities $\rho_{\alpha \alpha}-\rho_{\bar \alpha \bar \alpha}$
in the physical variable $x$. The prominent and extremely narrow resonance
structure (one cannot separate the neutrino and antineutrino resonances
here) is clearly visible.  The same spectrum is shown in the integration
variable $v$ in figure (\ref{fig:llv}). Now the sharp kink-like resonant
structure has broadened into a broad wave, which is easily represented
by the linearly discretized function in variable $v$ (grid points are
shown by the open circles). In this example we used just 400 grid points,
but essentially identical results were found with only 200 points.
Moreover, comparing the scales in figures~(\ref{fig:rplus}-\ref{fig:llx}),
one sees that the scale of variation of $L(x)$ is ten orders of magnitude
below unity. So, in order to follow the evolution of $L^{(\alpha)}$
accurately when using the original variables, one should be able to
integrate over distributions with an accuracy much better than ten
digits!  This seems totally impossible, and clearly indicates that
our use of variables $P^\pm$ is essential.

These examples prove the success of the key elements in our approach,
namely that by separating the small and large variables, and by
introducing the novel dynamical discretization of the momentum grid,
we can model the evolution of the ensemble of oscillating neutrinos
accurately.  Moreover, we can do so using small enough lattices such
that the code can be run even in a powerful modern workstation
(although a single run may then take as long as a couple of days).

\subsection{Regulating the sterile neutrino spectrum}

As one lets the system evolve further the resonance $x$ moves to
larger values while the integrated $L^{(\alpha)}$ decreases. This
continues until one reaches roughly the point when the main bulk of
the active neutrino distribution is getting through the resonance.
At this point one expects that the total asymmetry begins to oscillate
or grow, depending on the parameters. In our initial runs we did not
get that far however, but the program crashed due to an accumulation
of structures in sterile neutrino spectrum with momenta less than
$x_{res}T$.  How these structures arise is easy to understand.
Indeed, the active sector distributions are kept smooth away from
the resonances by diffusion induced by collisions with the background
particles. Sterile states interact only through the effective
interaction due to mixing however, and since this interaction
is neglibly small away from the resonances, the diffusion does not
work on sterile sector. As a result the moving resonance leaves
the sterile neutrino momentum distribution quite oscillatory for
$x < x_{res}$. While these developments in the sterile sector have
little effect on variables in the active sector (because active
and sterile states are essentially decoupled away from resonances),
they still pose a numerical problem: as the resonance moves forward
in $x$, the density of points reduces drastically in the region
$x\ll x_{res}$, until the point that the grid becomes too sparse
to model the oscillatory structures in the unsmoothed sterile
neutrino distribution. Eventually the amplitudes of these structures
overflow causing numerical instability and the solution breaks down.

The above explanation of the instability also suggests an obvious
cure for the problem. If one is not interested in studying the sterile
neutrino asymmetry spectrum, all we need to do is to add a small
regulatory interaction term to sterile neutrino sector, which will
smooth their distribution away from resonances, and yet does not
alter the evolution of the variables in the active sector. The latter
requirement can be met by a suitable form of the interaction term and
by making it small enough so that it does not change the physics in
the resonance region. The challenge is to do this so that the
regulatory term remains efficient enough to numerically stabilize the
system. To this end we have added the following ``sterile repopulation
terms" onto equation (\ref{steriili}):
\begin{eqnarray}
    R_{\nu_s}^+ &=&
      r_s \Gamma \left(n_{\nu_s}f_{eq}^+(x,\mu)
                     - \rho_{ss}^+  \right)
\label{rplusa} \\
    R_{\nu_s}^-  &=&
      r_s \Gamma \left( \rho_{aa}^{\rm init}
                     - \rho_{aa}^{-} - \rho_{ss}^- \right).
\label{regulators}
\end{eqnarray}
Here $\rho_{aa}^{\rm init}$ is the initial density matrix in the
active sector. We used the active sector interaction rate $\Gamma$
above, and let the parameter $r_s$ control the size of the regulatory
term.  The regulators (\ref{rplusa}-\ref{regulators}) differ from the
repopulation terms in the active sector in (\ref{aktiivi}) in that we
have added normalization factors to make sure that the integrals
over $R_{\nu_s}^\pm$ always vanish. (In this way we avoid spurious
sterile state equilibration even with large control parameter $r_s$.)

\begin{figure}
  \centering
  \includegraphics[height=7.5cm]{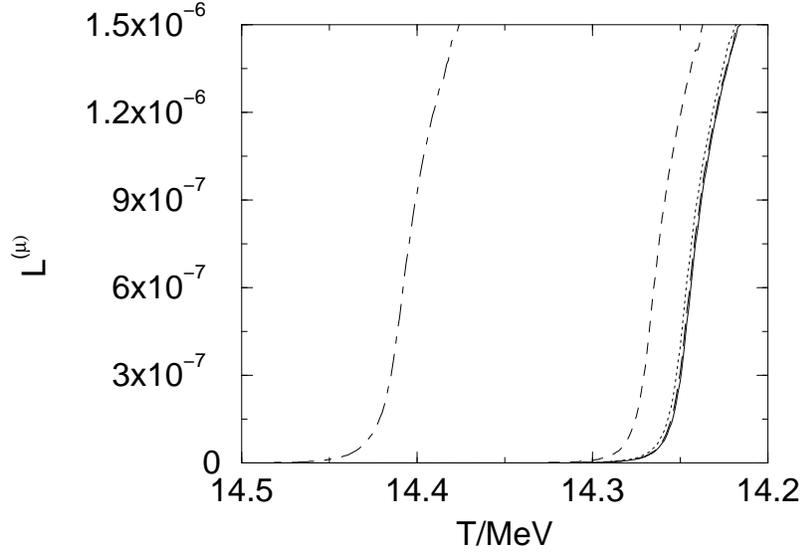}
  \caption{The stability of asymmetry evolution against variation of
    parameter $r_s$. We have used values $r_s =0.0001$ (solid line),
    0.001 (dashed line), 0.01 (dotted line) and 0.1 (dash-dotted line).
    Solutions for values $r_s=10^{-4}$ and $r_s=10^{-5}$ (long dashed line)
    are already
    indistinguishable in the figure. Oscillation parameters are
    $\delta m^2 = -0.1 \ev^2$ and $\sin^2 2 \theta_0 = 10^{-6.1}$.}
  \label{fig:regulator}
\end{figure}

We obviously need to show that there exists a range of values
for which the results converge and become independent of $r_s$, while
the regulator still continues to stabilize the solution.  That we do
reach this goal is evidenced in Fig.~(\ref{fig:regulator}), where we
show the evolution of the total asymmetry for $r_s=0.1$, 0.01, 0.001,
$10^{-4}$ and $10^{-5}$, for $\delta m^2 = -0.1 \ev^2$ and
$\sin^2 2 \theta_0 = 10^{-6.1}$.
Unsurprisingly, for large $r_s$ there is a significant
effect, but for small $r_s$ the results do converge to the point that
no visible difference can be seen between the two curves corresponding
to two smallest $r_s$'s.  Let us stress that the mixing parameters
were here deliberately chosen such as to show a particularly large
sensitivity for the regulator.  For many other cases that we studied
it would have been difficult to show any effect at all; for example
the results displayed in figure~(\ref{fig:stable}) showed no
appreciable dependence on the regulator up to $r_s = 10$.

We have also studied the stability of our results against changes in
various numerical error handling parameters in our code. The stepwise
error tolerance for example can be relaxed by an order of magnitude
from what we actually employed, before any visible deviation from the
solution can be seen over the entire integration range. This is true also
for our examples corresponding to the oscillatory solutions in or
near the chaotic region.  Of course, the numerical accuracy would be
eventually compromised for parameters for which the asymmetry undergoes
very large number of oscillations before settling to the growing
curve, but as in~\cite{eks1}, that does not affect our main conclusions.
Also cut off parameters $x_{min}$ and $x_{max}$ and the tilt $\alpha$
which controls the density of points at the resonance can be varied
within reasonable limits without any observable effects on the results.
We therefore are confident that our observed pattern of asymmetry
oscillations is indeed physical, and not of numerical origin.

\subsection{Argument for chaoticity in $L$-growth}

Having introduced a method that can handle the evolution of momentum
dependent density matrices and in particular the asymmetry, and having
proven its numerical stability and accuracy, we now pursue the question
of the ``chaotic" behaviour of the final sign of the integrated asymmetry.
Note that while the use of word chaos is customary in this context, we
strictly speaking mean only sensitivity of the final sign of the asymmetry
on initial seed asymmetry and on oscillation parameters\footnote{
In fact the system {\em does} exhibit true chaoticity at a certain
level, but the related information loss is very small in
practice~\cite{Steen}.}.

In the context of our earlier work using the momentum averaged
equations~\cite{eks1,Sorri}, we explained what was the physical origin
for the sign sensitivity. Quite simply, changes in the boundary conditions
(initial baryon asymmetry and the oscillation parameters) change the
direction of the motion of the system, and the topography of the
attractor solutions in the phase space at the onset of resonance. Roughly
speaking, in the momentum averaged case one can draw an analogy between the
system and a somewhat sticky ball initially rolling down a single valley
floor that later branches to two, with a low maximum in between, at the
resonance point. Changing baryon asymmetry would then be analogous to
changing the speed and direction of the ball at the branching point and
changing oscillation parameters to that of changing the shape of the
valleys themselves. Sign sensitivity arises for those initial conditions
and valley forms for which the ball makes many oscillations between the
two degenerate valleys before stickiness (damping) eventually wins and
makes it to settle into either one of them.

In the momentum dependent case these structures become smeared out, and
to draw a mechanical analogy, one should rather think, instead of a ball,
of a string of varying size beads (largest ones in the middle, corresponding
to the peak in the thermal distribution) coupled by a very elastic cord,
wiggling down the valleys described above.  While this is obviously a
much more complicated system with qualitatively new features, one would
still expect to see oscillations in the weighted average position of beads
at least when the largest ones roll past the branching point (in $L$ when
the maximum of the distribution crosses the resonance). Moreover, if the
average position of the string ($L$) makes many oscillations over the
central hill before stickiness (damping) wins, one again should expect
that the final valley that the string settles in (sign of $L$) strongly
depends on initial and boundary conditions. While this analog is certainly
bears only a crude resemblance to the true system, it should help convince
the reader of the fact that there is nothing mysterious about the sign
sensitivity in the $L$-evolution, but that it is rather something to be
expected.

\begin{figure}
    \centering
    \includegraphics[height=7.5cm]{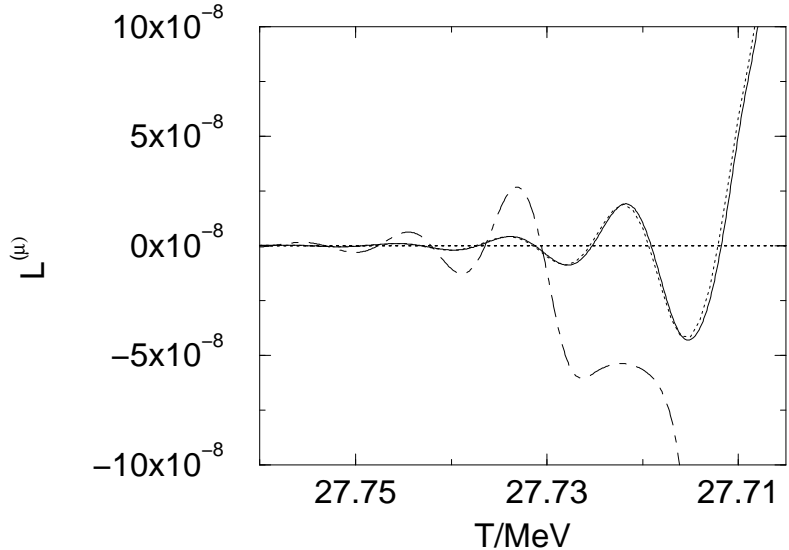}
    \caption{Neutrino asymmetry oscillations are shown near the boundary
    of "chaotic" region using same $\delta m^2 = -10 \ev^2$ and three
    slightly different mixing angles $\sin^2 2 \theta_0 = 10^{-6.1}$
    (solid line) $\sin^2 2 \theta_0 = 10^{-6.099}$ (dotted line) and
    $\sin^2 2 \theta_0 = 10^{-6.09}$ (dash-dotted line).}
    \label{fig:osc}
\end{figure}

Obviously, all we need to do to prove the existence of chaoticity, is
to find examples of almost identical sets of oscillation parameters for
which the final asymmetry does show significantly different oscillation
pattern, and leads to a different final sign. We show such a case in
figure (\ref{fig:osc}). All three curves in the figure correspond to
$\delta m^2 = -10 \ev^2$ but have slightly different vacuum mixing
angles. The two curves which behave almost alike, have mixing angles
$\sin^2 2 \theta_0 = 10^{-6.1}$ (solid curve) and $\sin^2 2 \theta_0
= 10^{-6.099}$ (dotted curve) respectively. The dash-dotted curve
with a very different behaviour corresponds to $\sin^2 2 \theta_0
= 10^{-6.09}$. So, if such mixing were ever observed and the mixing
angle was measured to a precision better than a tenth of a per cent,
the final sign of the lepton asymmetry and its effect on nucleosynthesis
could be predicted from our computation.  However, if the mixing angle
was measured with less than about one per cent accuracy, we would not
be able to say what the final asymmetry will be, and therefore what is
the precise BBN prediction for helium abundance.

\begin{figure}
  \centering
  \includegraphics[height=7.5cm]{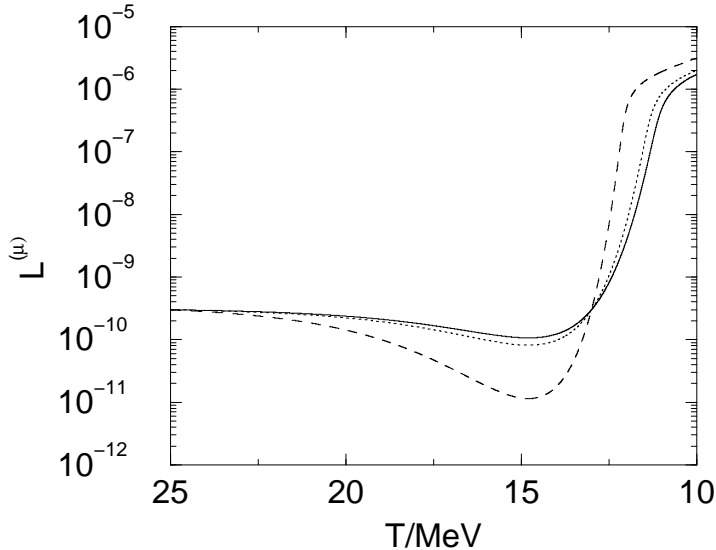}
  \caption{Asymmetry evolution in the stable region of oscillation
  parameters.  All curves corresponds to mass splitting $\delta m^2
  = -0.1\;\rm eV^2$ and the mixing angles are $\sin^22\theta_0 =
  10^{-9}$ (solid curve), $\sin^2 2\theta_0 = 10^{-8.9}$ (dotted curve)
  and $\sin^2 2\theta_0 = 10^{-8.5}$ (dashed curve).}
  \label{fig:stable}
\end{figure}

The sign sensitivity does not occur for all oscillation parameters,
however. In the figure (\ref{fig:stable}) we show results from ``stable"
region of parameters (same mass difference as above, but much smaller
mixing angle). In this case, while visible differences in the evolution
become visible when mixing angle is changed by a factor or three, the
asymmetry never becomes oscillatory, and the final sign is determined
by the initial sign of the asymmetry.

Our first example was chosen from close to the edge of the chaotic
region, and just as making $\sin^22\theta_0$ smaller made system more
stable, increasing it has the opposite effect. The number of oscillations
and also degree of sensitivity of the final sign to the oscillation
parameters increases very rapidly with $\sin^22\theta_0$. Nevertheless,
the degree of sensitivity to parameters is weaker and thus the chaotic
region smaller than what was found in the momentum averaged
treatment~\cite{eks1}. Quantitatively our new results agree roughly
with the boundaries of the chaotic region inferred in~\cite{fdb}, but
since the numerical work is, even with the improvements we have made,
very time consuming, we will not attempt a precise mapping of the
chaotic region here.

\begin{figure}
    \centering
    \includegraphics[height=7.5cm]{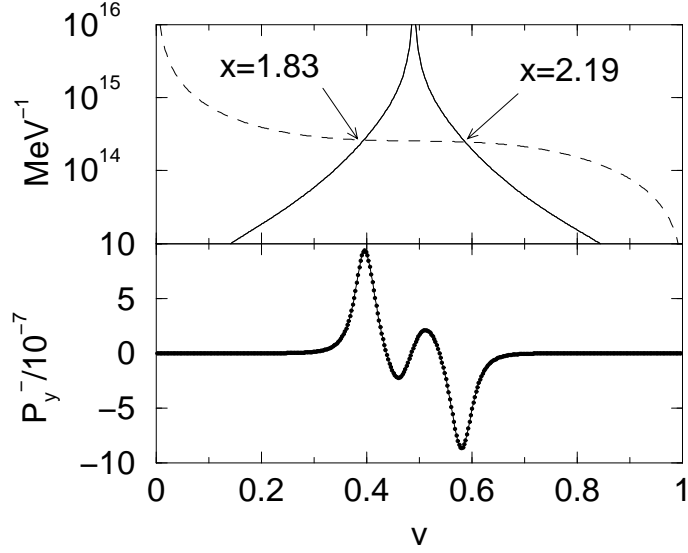}
    \caption{Lower panel: Variable $P_y^-$ as a function of $v$ at
            temperature $T=27.8 \mev$ for oscillation parameters
            $\delta m^2 = -10 \ev^2$, $\sin^2 2\theta_0 = 10^{-6.099}$.
            Upper panel: The coherence length (dashed line) and
            $l_{\rm m}/4$ (solid line) for the same parameters.
            Physical momentum at the resonance is $x_{\rm res} = 1.88$.}
    \label{fig:py}
\end{figure}

Let us finally to try to understand a little more deeply what precisely
drives the oscillations of the total asymmetry in the momentum dependent
case.  A clue can be found from Figure~(\ref{fig:py}) which shows the
variable $P_y^-$ in the region where the asymmetry oscillates significantly.
(At earlier times, when the total $L$ does not oscillate, the structure
of the $P_y^-$-spectrum is similar to that of $\rho_{\alpha\alpha}^-$ shown
in Fig.\ (\ref{fig:llv}).)  The two new peaks seen outside the resonance
grow slowly in time, while their sign oscillates rapidly. Because these
structures are asymmetric around the resonance (in the physical variable
$x$), the integral over $P_y^\pm(x)$ is nonzero and oscillates. As a
result also $L$, which is solved from (\ref{L-equation2}), becomes
oscillatory.

What causes these structures, is a delicate interplay between the
coherence length and the matter oscillation length. The former is simply
given by the inverse of the damping rate $\ell_{\rm coh} = 1/D$ and the
latter is
\begin{equation}
\ell_{\rm m} = \frac{l_{\rm vac}}{\sqrt{1-2 \chi\cos2\theta_0 + \chi^2}}
\end{equation}
where $l_{\rm vac} \equiv 4\pi p/|\delta m^2|$ and
$\chi \equiv 2p|V|/|\delta m^2|$ where $|V|$ is the magnitude of
the matter contribution to the $V_z$ in equation (\ref{vzeta}).
Near the resonance the matter oscillation length is very long, and in
particular at high temperatures it exceeds the coherence
length by a large margin.  In such a situation the resonance is strongly
{\em overdamped}, and despite the maximal mixing angle oscillations
do not have time to develop because the state is continuously projected
to the active direction. This in part explains why the resonant features
have so modest amplitudes at early times (cf.\ Fig.\ (\ref{fig:llx})).

In the upper panel of the figure (\ref{fig:py}) we have plotted
the coherence length and $\ell_{\rm m}/4$, which
corresponds to the distance over which $P_y^\pm$ first rises to
maximum due to oscillation. It is clear that the new structures
correspond to the case when the matter oscillations first time
break the overdamping situation.
Further away from the resonance the damping no more can block the
oscillations, but since the matter mixing angle decreases very fast
away from the resonance, the amplitude of the oscillation features
dies off quickly. As the temperature decreases, the new structures move
away from the resonance, and they also become broader (one can show
that their separation scales like $\delta x \sim T^{-3}$). This is why
their amplitude slowly increases in time (with decreasing $T$) causing
larger and larger amplitude oscillations in the total asymmetry
$L$.  The lesson to bring home from these observations is that the
dynamics of $L$-growth and oscillations is very dependent on the
detailed fine structure of the various components of the density
matrix.  Such fine details on the hand can only be studied in a
numerical approach using both separated small and large variables
($P_i^\pm$) and dynamically adjusting momentum grids.

\section{Conclusions}

Active-sterile mixing, while constrained by the present solar
and atmospheric data, are nevertheless required if one wishes to
incorporate the LSND-neutrino anomaly into the neutrino models.
Moreover, the active-sterile neutrino mixing would have a rich
phenomenology in the early universe, where it provides a unique
theoretical challenge in the form of a system for which the
quantum effects play an essential role in the kinetic equations.

Here we have studied the phenomenon of neutrino asymmetry growth in
the early universe as a result of active sterile neutrino oscillations.
Our main new contribution is the introduction of a novel method of
discretizing the momentum variable such that the sharply pronounced
structures near the resonances, which are here shown to drive the
oscillation phenomena, can be treated numerically accurately.
The only concession to rigor we made along the way to solution of
the problem, was adding collision terms to regulate the sterile
neutrino spectrum away from the resonances. However, we demonstrated
that our regulators had no effect on the active neutrino evolution,
and hence for the results presented here. Nevertheless, our treatment
is obviously not adequate if the precise form of the sterile neutrino
spectrum is important.  This might be the case for the models where
sterile neutrinos are invoked to provide a non-thermal component for
the dark matter~\cite{nonThDM}.

We have demonstrated that even tiny changes in the oscillation
parameters may drastically alter the oscillation pattern, and even
change the sign of the final asymmetry.  This behaviour does not
occur for all oscillation parameters, but instead the dependence of
the sign of $L$ on oscillation parameters is weak in the ``stable"
region (in general small $\sin^22\theta_0$ and large negative
$\delta m^2$) and strong in the chaotic region (in general large
$\sin^22\theta_0$ and small negative $\delta m^2$). These results are in
qualitative agreement with our earlier findings, which were based on
a momentum averaged treatment~\cite{eks1}.  Obviously, if
active-sterile mixing were to be observed with parameters residing
in the chaotic region, it would be, due to inevitable errors in the
experimentally measured parameters, impossible to reliably determine
the sign of the neutrino asymmetry created by that mixing in the
early universe.  Moreover, since large neutrino asymmetries (and
electron neutrino asymmetry in particular) directly affect, in a
$sign(L)$-dependent way, the weak interaction rates that
determine proton-to-neutron ratio in the early universe, this
indeterminacy would undermine our ability to accurately compute the
BBN prediction for light element abundances for chaotic oscillation
parameters.

These conclusions hold in the spatially homogeneous calculation.
However, the sensitivity of $L$-growth on initial conditions also
lends to a speculation that in reality the initially inhomogeneous
seed asymmetry might give rise to an inhomogeneous texture of domains
of very large lepton asymmetries with oscillating sign. This phenomenon
has been shown to occur in a one-dimensional, momentum averaged
model~\cite{eks2}, and could plausibly occur in a realistic three
dimensional world.  The implications of such a scenario for SBBN would
obviously be very different, including the possibility of very efficient
equilibration of sterile neutrinos via MSW-effect within the domain
boundaries~\cite{shiFuDo,eks2}. If this was the case, then the asymmetry
growth mechanism would lead to even {\em stronger} bounds on mixing than
what is displayed in equations~(\ref{cosmofluxes}).  It is beyond the scope
of this work (and the reach of the present computers) to study this
phenomenon quantitatively in the momentum dependent case, however.

Let us finally comment on the effects of large homogeneous lepton
asymmetries for cosmic microwave background radiation (CMBR). The
possibility of measuring $L$ by the Planck satellite data has been
considered for example in~\cite{Kinney:1999pd} and
in~\cite{CMBconstraints}.  However, the only effect of $L$ on CMBR
comes through the associated fluctuations in the energy density,
and correspondingly only the total asymmetry has relevance.
In the oscillation scenarios, such as discussed in this paper,  the
total asymmetry (here the sum of the sterile and active sector
asymmetries) is conserved however, and hence the oscillation-induced
asymmetries, in contrast to the situation with SBBN, would have
no  direct effect on CMBR.

\acknowledgments

We thank Kari Enqvist for useful conversations and a collaboration
at earlier stages of this project. We also thank Steen Hannestad
for discussions on the effects of $L$ on CMBR. AS wishes to thank
Nordita for hospitality during several visits in the course of
completing this project.

%
%

\end{document}